# Neural CRC Prediction for 5G NR URLLC

Prashanth Murthy

Dell Technologies Inc., Round Rock TX 78664, USA
prashanth.murthy@dell.com

**Abstract.** We propose a neural cyclic redundancy check (CRC) predictor for the 5G New Radio (5G NR) physical uplink shared channel (PUSCH) that enables early link-adaptation decisions for Ultra-Reliable Low-Latency Communications (URLLC). The predictor combines a lightweight convolutional neural network (CNN) with a fixed front-end that extracts multi-scale time-frequency energy features from the received signal and least-squares channel estimates. We investigate two complementary front-end realizations – a wavelet scattering front-end built from fixed Gabor filters, and an FFT-based scattering front-end that applies bandpass masks in the frequency domain with geometric scale spacing. Drawing on neural-receiver design principles, the predictor estimates the post-decoding CRC outcome directly from the received resource grid, bypassing the conventional equalization and decoding chain. We further introduce the modulation and coding scheme (MCS) index as an auxiliary conditioning input that adapts the decision boundary to the operating code rate. Experiments on a multi-MCS 5G NR PUSCH dataset show that the hybrid scattering predictors substantially outperform a pure CNN baseline, with MCS conditioning further improving reliability across varying channel conditions. Both front-ends are compatible with GPU-accelerated inference and are lightweight enough to be deployed within a real-time baseband pipeline. We further demonstrate an evidential deep learning extension that quantifies epistemic uncertainty in a single forward pass using a conservative decision rule.

## 1 Introduction

Neural networks have increasingly been adopted in wireless communication systems due to their ability to approximate complex nonlinear functions and to learn representations directly from data. In cellular baseband processing, receiver functions such as channel estimation, equalization, and symbol detection are traditionally implemented as separate algorithmic blocks designed under simplifying assumptions about the channel and noise statistics. Recent research has explored the concept of a neural receiver, in which several of these stages are replaced or augmented by a neural network that operates directly on the received I/Q samples and learns to recover the transmitted information.

Conventional receivers for the 5G NR PUSCH typically employ variants of minimum mean-squared error (MMSE) equalization followed by demodulation and LDPC decoding, with a CRC check appended at the end to verify the correctness of the decoded transport block (TB). While these receivers are computationally efficient and widely deployed, they rely on assumptions such as Gaussian interference, perfect or near-perfect channel estimation, and simplified channel models. In practice, wireless channels are affected by multipath fading, Doppler shift, cell-edge interference, and hardware impairments that can violate these assumptions and degrade receiver performance. Because the CRC check is the final step of this chain, any failure caused by these real-world impairments is only detectable after the entire equalization, demodulation, and decoding pipeline has been completed.

Neural receivers seek to overcome the shortcomings of conventional receiver designs by learning both the wireless channel behavior and the optimal detection strategy directly from data. This concept can be extended to CRC prediction as well; instead of using the CRC as a final consistency check after the full decoding process, a neural model can directly infer the decoding outcome from the received signal itself, effectively skipping intermediate stages that are more prone to breakdown under challenging channel conditions.

While neural receivers focus on symbol detection and soft-bit generation, the same data-driven processing of the received resource grid can be repurposed for a different and operationally valuable task, predicting the outcome of the CRC before the transport block (TB) is fully decoded. In this work we draw inspiration from neural-receiver architectures, specifically their use of the received uplink I/Q samples together with least-squares (LS) channel estimates as network inputs and apply this processing paradigm to early CRC prediction for URLLC.

### 1.1 Motivation

Ultra-Reliable Low-Latency Communications (URLLC) is one of the three primary 5G service categories, targeting end-to-end latencies below 1 ms and reliability above 99.999%. It underpins mission-critical applications such as industrial automation, autonomous vehicles, remote surgery, smart grids, and real-time robotics, where even minor delays or sporadic data loss are unacceptable.

To meet these constraints, 3GPP introduced several radio-resource enhancements that generally trade spectral efficiency for reduced latency. Two key enablers, and their limitations, motivate this work:

- Mini-slot scheduling. A mini-slot spans fewer OFDM symbols than a normal slot (e.g., 2, 4, or 7 symbols instead of 14), reducing transmission duration. However, when PUSCH is scheduled toward the end of a slot, the gNB processing time constrains the hybrid automatic repeat request (HARQ) feedback and the uplink grant for the next slot. This adds approximately one slot of latency, and any retransmission triggered by a decoding failure further increases both latency and jitter, breaking the deterministic packet cadence that URLLC applications require.

- Configured-grant repetitions. To avoid the HARQ round-trip-time (RTT) delay, the same TB may be transmitted blindly 2, 4, or 8 times using the same HARQ process and redundancy version. Repetitions improve reliability but waste air-interface resources and, in the absence of link adaptation, may not help under specific fading realizations.

Although these two mechanisms are complementary to each other, a solution that reduces the HARQ RTT and avoids unnecessary repetitions would be beneficial to conserve air interface resources while providing the required quality of service. Recent system-level studies have shown that URLLC performance is fundamentally constrained by tightly coupled protocol and processing delays, particularly those associated with scheduling and feedback timing [1]. This is precisely the gap addressed by early CRC prediction. If the decoding outcome can be predicted reliably and quickly, a retransmission can be scheduled at the earliest opportunity (e.g., the very next slot) through normal link adaptation, restoring deterministic latency without resorting to blind repetition.

Fig. 1 illustrates the bottleneck using the commonly deployed 5G NR TDD slot pattern DDDSUUDDDD (10 slots per 5 ms half-frame at 30 kHz SCS, μ=1). A PUSCH mini-slot occupying symbols 3–9 of the first uplink slot (slot #4 in the pattern) leaves only symbols 10–13 or, approximately 142 μs before the slot boundary. Full LDPC decoding of a TB using the full 100 MHz bandwidth across 7 OFDM symbols cannot typically complete in this window; the gNB therefore has no decode result before the slot ends and cannot schedule a retransmission for the immediately following uplink slot (slot #5). The retransmission is consequently deferred to the next period's uplink slots, adding a full 5 ms of HARQ RTT. A neural CRC predictor that completes within the remaining symbols of slot #4 eliminates this delay since its binary pass/fail estimate is available before the slot boundary, enabling the gNB to schedule a retransmission in slot #5 and reducing the effective HARQ RTT from 5 ms to 0.5 ms. The same benefit applies to eMBB on a wider allocation, but the impact is most pronounced for URLLC mini-slots where every slot boundary is a hard latency cliff.

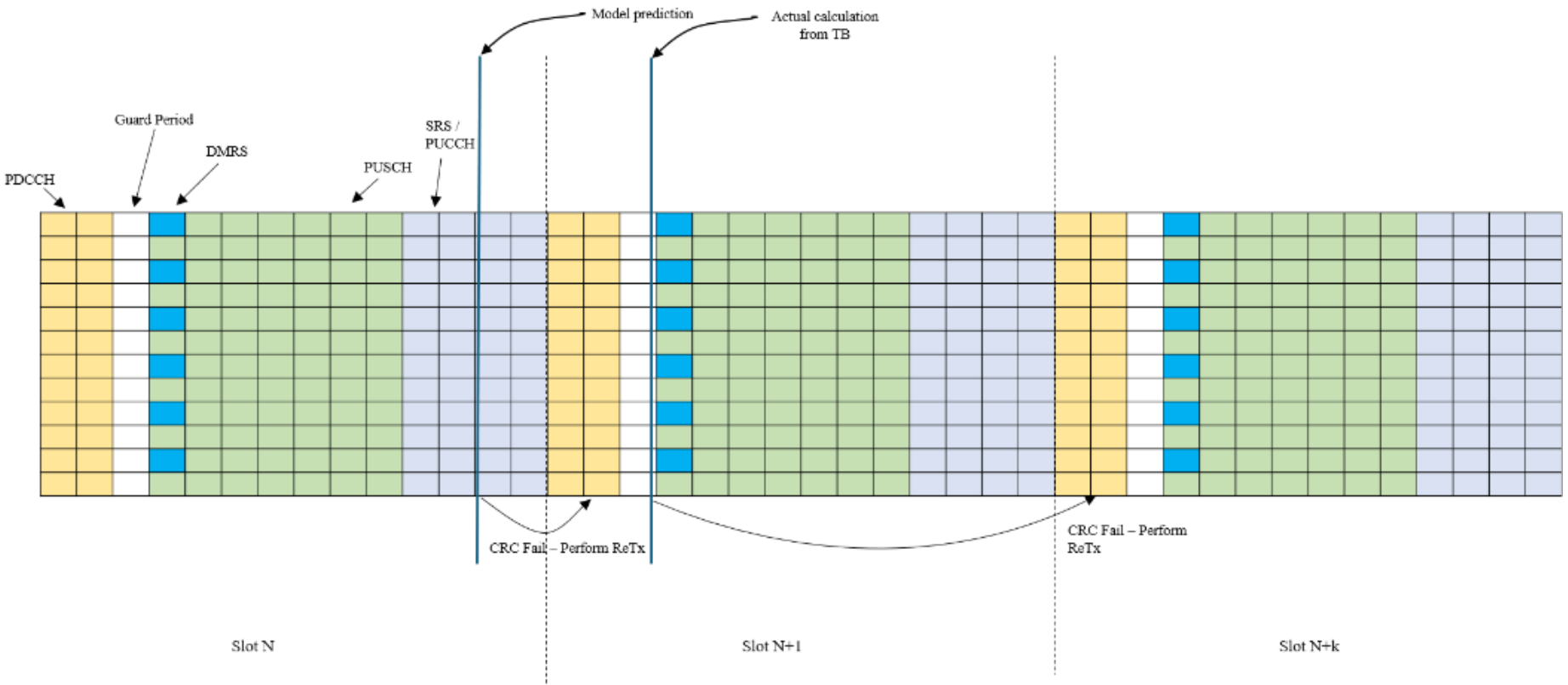

**Fig. 1.** CRC prediction vs calculation from decoded TB

### 1.2 Related Work

Early work demonstrated that neural networks could perform symbol detection directly from received signals, often outperforming traditional algorithms under challenging channel conditions. Examples include the fully convolutional DeepRx receiver [2], which jointly performs equalization and detection for OFDM systems, and model-driven approaches such as OAMP-Net [3], which embed neural components into classical iterative detectors. Hybrid architectures that combine signal-processing insight with learnable components have also been studied to improve robustness and generalization [4].

More recent work has investigated the practical integration of neural receivers into real-world systems [5, 6], including implementations within GPU-accelerated RAN frameworks such as NVIDIA Aerial, demonstrating the feasibility of deploying neural detectors in 5G NR baseband pipelines. Low-power alternatives based on spiking neural networks have likewise been proposed [7]. These efforts established that processing the received I/Q grid (optionally conditioned on LS channel estimates) with a compact neural network is both feasible and effective – the design principle we adopt for the CRC prediction front-end.

The feature-extraction stage of our predictor is grounded in the scattering transform framework introduced by Bruna and Mallat [8], which builds translation-invariant, stable representations by cascading wavelet convolutions with modulus nonlinearities. We retain only the first-order scattering coefficients and study two implementations – a Gabor-filter realization in the spatial domain, and an FFT-domain realization in the spirit of fast scattering computations [9]. To our knowledge, the use of scattering front-ends for CRC prediction, and the comparison of Gabor versus FFT realizations under MCS conditioning, has not been previously studied.

### 1.3 Contributions

The main contributions of this work is summarized as follows:

- We formulate early CRC prediction for 5G NR PUSCH as a binary classification problem operating on the received I/Q grid and LS channel estimates, and we connect the two decision errors – missed detections and false alarms, directly to the URLLC retransmission trade-off.
- We propose a hybrid predictor that combines a compact CNN classifier with a fixed scattering front-end, with two realizations – a Gabor wavelet implementation using spatial-domain depthwise convolutions, and an FFT-based implementation using frequency-domain bandpass masks with geometric scale spacing. We analyze when each is preferable in terms of accuracy, latency, and accelerator mapping.
- We introduce MCS conditioning as an auxiliary scalar input and show that it resolves the high false-alarm rate of scattering models operating across multiple code rates, improving accuracy and halving the false-alarm rate.
- We demonstrate that both front-ends are compatible with TensorRT and execute within the PUSCH slot processing budget on an edge-class GPU, with the FFT front-end achieving the lowest measured TensorRT kernel time.

# 2 System Model

## 2.1 Uplink Signal Model

We consider uplink transmission of the 5G NR PUSCH over a frequency-selective fading channel. Let $x[k,l]$ denote the transmitted complex-valued symbol on subcarrier $k$ and OFDM symbol $l$, drawn from a modulation constellation such as QPSK or QAM. The base station is equipped with $N_r$ receive antennas; the received signal on antenna $r$ is

$$y_r[k,l] = h_r[k,l]\,x[k,l] + n_r[k,l], \qquad r = 1,\dots,N_r \tag{1}$$

where $h_r[k,l]$ is the channel frequency response from the UE to antenna $r$ on resource element $(k,l)$ and $n_r[k,l]$ is additive complex Gaussian noise. In the experiments reported here $N_r = 2$. In practical wireless environments, the received signal may also be affected by multipath fading, Doppler shift, cell-edge interference and hardware impairments. These effects make accurate channel estimation and equalization challenging for traditional receivers that rely on simplified channel assumptions.

## 2.2 Conventional Receiver Processing

In a conventional receiver, the transmitted symbols are recovered through a sequence of signal processing blocks that typically include channel estimation, equalization, and demodulation. A widely used approach is the linear minimum mean-square error receiver, which computes an estimate of the transmitted symbol as

$$\hat{x}[k,l] = \frac{\mathbf{h}^H[k,l]}{\|\,\mathbf{h}[k,l]\,\|^2 + \sigma_n^2}\,\mathbf{y}[k,l] \tag{2}$$

where $\mathbf{h}[k,l] = [h_1[k,l],\dots,h_{N_r}[k,l]]^T$ is the stacked channel vector, $\mathbf{y}[k,l]$ ] is the corresponding received vector, and $\sigma_n^2$ denotes the noise variance. The equalized symbols are demodulated to log-likelihood ratios (LLRs), de-rate-matched, and passed to the LDPC decoder. A CRC is then computed over the decoded TB to verify correctness. Let $c \in \{0,1\}$ denote the resulting CRC outcome, with $c = 1$ indicating a successfully decoded (CRC-pass) TB and $c = 0$ a failed (CRC-fail) TB

$$c = \mathrm{CRC}\big(\mathrm{Decode}(\hat{x})\big) \tag{3}$$

The outcome $c$ is only available after the full decoding chain has been completed, which is precisely the latency that URLLC cannot afford when a retransmission is required.

## 2.3 CRC Prediction Formulation

The neural CRC predictor estimates the probability that the TB will pass the CRC directly from the received signal and channel estimates, bypassing equalization, demodulation, and decoding, as illustrated in Fig. 2 Following the neural-receiver input convention [2, 5], the predictor consumes the received data symbols $y$ and the LS channel estimates $\hat{h}_{\mathrm{LS}}$ obtained from the demodulation reference signal (DMRS). Since DMRS occupies only a subset of resource elements, the LS estimates are expanded to the full resource grid prior to feature extraction.

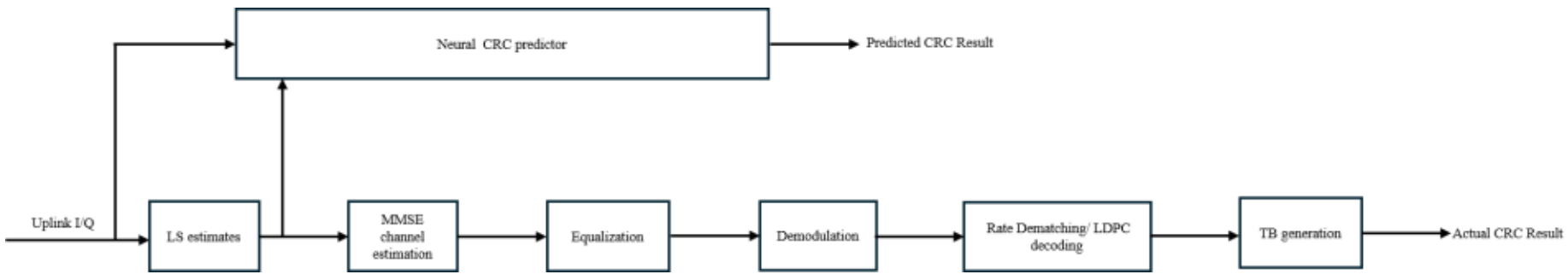

**Fig. 2.** CRC prediction in the PUSCH pipeline

Let $F_\theta(\cdot)$ denote a neural network parameterized by weights $\theta$. The predictor estimates the posterior probability of a CRC pass as

$$\hat{p} = P\left(c = 1 \mid y, \hat{h}_{\mathrm{LS}}\right) = \sigma\left(F_\theta(u)\right) \tag{4}$$

where $\sigma(\cdot)$ is the logistic sigmoid and $u$ is the joint input representation defined in Sect. 3.

The network is trained against the true CRC labels obtained from the conventional decoder using the binary cross-entropy (BCE) loss

$$\mathcal{L} = -\sum_i [\, c_i \log\hat{p}_i + (1 - c_i)\log(1 - \hat{p}_i)] \tag{5}$$

Unlike neural receivers that regress LMMSE-derived soft targets, the CRC predictor is supervised by the ground-truth decoding outcome, making it a calibrated binary classifier. At inference, a decision is obtained by thresholding the predicted probability,

$$\hat{c} = \mathbb{1}[\, \hat{p} \geq \tau\, ] \tag{6}$$

where the threshold $\tau$ (default 0.5) can be tuned to bias the predictor toward catching failures, as discussed in Sect. 4.

### 2.4 Operational Objective and Error Metrics

The predictor's value is determined by how its two error types map onto the URLLC retransmission decision. Adopting the convention $c = 0$ (CRC fail, negative) and $c = 1$ (CRC pass, positive), Table 1 summarizes the actions taken at prediction time $t$ versus the eventual outcome at decode time $t + \Delta$. Two rates are operationally critical, defined directly in terms of the prediction outcome:

- Missed Detection Rate (MDR): the fraction of truly failed TBs ($c = 0$) that the predictor labels as a pass. This is a silent error – no early retransmission is scheduled, and the latency-saving opportunity is lost. MDR is the safety-critical metric for URLLC.

$$\mathrm{MDR} = \frac{|\{i: c_i = 0,\ \hat{c}_i = 1\}|}{|\{i: c_i = 0\}|} \tag{7}$$

- False Alarm Rate (FAR): the fraction of truly passed TBs $c = 1$ that the predictor labels as a failure, triggering an unnecessary retransmission. A false alarm wastes resources, but far less than blind repetition, which retransmits every TB 2–8×.

$$\mathrm{FAR} = \frac{|\{i: c_i = 1,\ \hat{c}_i = 0\}|}{|\{i: c_i = 1\}|} \tag{8}$$

This asymmetry frames the design objective as – minimize MDR to preserve URLLC determinism, while keeping FAR low enough that the resource savings over blind repetition are retained.

**Table 1.** Action table mapping predicted CRC at time $t$ to action at decode time $t + \Delta$

| Scenario | **Relative time** | **CRC Observation** | **Action** |
|---|---|---|---|
| 1 | t | Predicted Negative | MAC performs LA and fast ReTx |
| | t + Δ | Actual Negative | No further action |
| 2 | t | Predicted Negative | MAC performs LA and fast ReTx |
| | t + Δ | Actual Positive | MAC schedules new Tx |
| 3 | t | Predicted Positive | No action |
| | t + Δ | Actual Negative | MAC performs LA and normal ReTx |
| 4 | t | Predicted Positive | No action |
| | t + Δ | Actual Positive | MAC schedules new Tx |

# 3 Proposed Hybrid Neural CRC Predictor

## 3.1 Architecture Overview

The proposed architecture is illustrated in Fig. 3. The receiver takes as input the received uplink data symbols $y$, represented as complex-valued I/Q samples on the resource grid, and the least-squares channel estimates $\hat{h}_{\mathrm{LS}}$ obtained from the demodulation reference signals (DMRS). Following the approach proposed in DeepRx [2], the channel estimates are used as conditioning inputs to the neural detector and along with MCS. Since DMRS symbols occupy only a subset of the resource elements, the LS channel estimates are expanded to the full resource grid prior to feature extraction.

The received data symbols $y$ and the expanded channel estimates $\hat{h}_{\mathrm{LS}}$ are concatenated along the feature dimension to form a joint representation

$$u = \mathrm{concat}(y, \hat{h}_{\mathrm{LS}}) \tag{9}$$

The LS channel estimates encode both the channel magnitude and, implicitly, the per-subcarrier SNR – since the estimation noise scales inversely with SNR – providing the predictor with an implicit measure of link quality without requiring an explicit SNR computation. The tensor $u$ is processed by a fixed scattering front-end $\mathcal{W}(\cdot)$ that extracts first-order, multi-scale, time-frequency energy features. The front-end is realized either with Gabor filters or with FFT-domain bandpass masks – both produce a phase-invariant feature tensor that is passed to the CNN classifier. When MCS conditioning is enabled, the MCS index is embedded and fused with the pooled CNN features before the final sigmoid (Sect. 3.4).

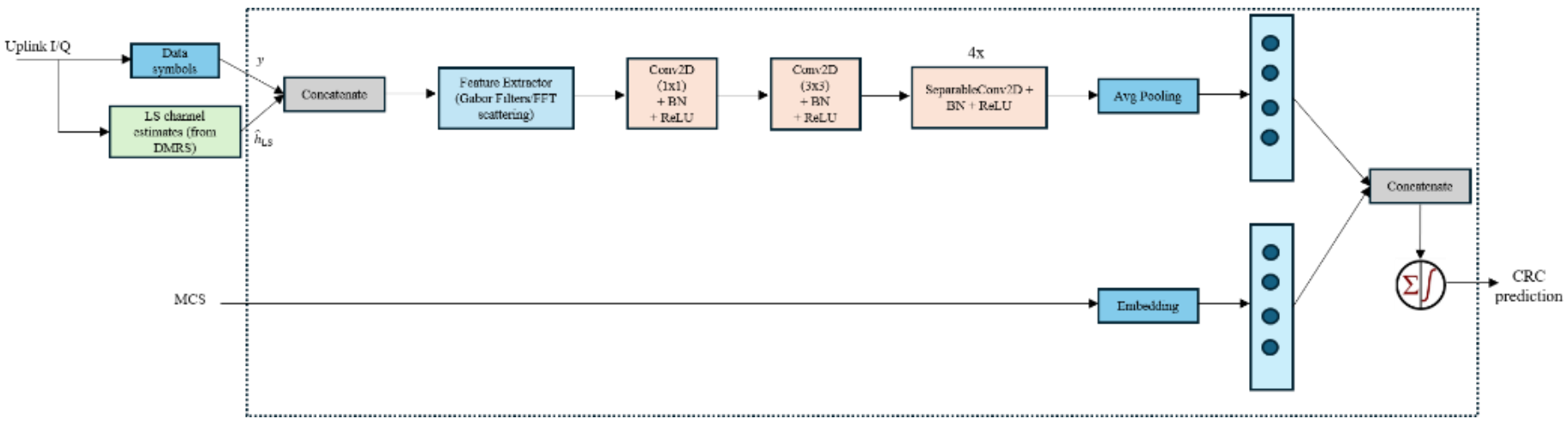

**Fig. 3.** Hybrid CRC predictor

### 3.2 Gabor Filter Front-end

The input tensor $u$ is processed using a bank of fixed Gabor filters that approximate the first-order wavelet scattering representation and are implemented as depthwise convolution layers.

Let $g_{j,\theta}$ denote a complex Gabor filter parameterized by scale $j$ and orientation $\theta$. The response of the filter bank is computed by convolving the input tensor with the real and imaginary components of the filter

$$u_r = u * g_{j,\theta}^{(r)}, \qquad u_i = u * g_{j,\theta}^{(i)}, \tag{10}$$

where $*$ denotes convolution.

To obtain a phase-invariant representation, the magnitude of the complex response is computed using a modulus operation

$$z_{j,\theta} = \sqrt{u_r^2 + u_i^2} \tag{11}$$

This operation extracts the local energy of the filtered signal and produces a set of feature maps capturing multi-scale variations in the received signal and channel estimates. Stacking the responses over all scales and orientations yields the scattering feature tensor

$$z = \mathcal{W}(u) \tag{12}$$

In the proposed implementation, the filter bank uses $J = 2$ scales and $L = 4$ orientations ($\theta \in \{0, \pi/4, \pi/2, 3\pi/4\}$), giving $J \cdot L = 8$ first-order coefficients per input channel. The scales are spaced linearly in the Gabor envelope width $\sigma$ and wavelength $\lambda$. Each $g_{j,\theta}$ is a zero-mean, energy-normalized $5 \times 5$ kernel realized as a non-depthwise convolution layer with depth multiplier $J \cdot L$. The modulus in (9) provides invariance to the residual phase rotations left by imperfect equalization, which vary with Doppler and delay spread, making the energy features robust descriptors of channel-induced distortion.

### 3.3 FFT-based Scattering Front-end

The same first order scattering features can be computed in the frequency domain by exploiting the convolution theorem, which states that convolution with $g_{j,\theta}$ is equivalent to multiplication by its transfer function $M_{j,\theta}$ in the Fourier domain. Let $\mathcal{F}$ and $\mathcal{F}^{-1}$ denote the two-dimensional discrete Fourier transform and its inverse. The input is first transformed to the frequency domain,

$$U = \mathcal{F}(u) \tag{13}$$

and a bank of bandpass filters is applied as element-wise multiplication with frequency-domain masks $M_{j,\theta}$,

$$V_{j,\theta} = U \odot M_{j,\theta} \tag{14}$$

where $\odot$ denotes the Hadamard product. The filtered spectrum is mapped back to the spatial domain, and the modulus yields the phase-invariant scattering coefficients

$$z_{j,\theta} = \left|\mathcal{F}^{-1}\left(V_{j,\theta}\right)\right| \tag{15}$$

so that the complete FFT-based scattering representation is

$$z = \mathcal{W}_{\mathrm{FFT}}(u) \tag{16}$$

Equation (13)-(16) are the frequency-domain counterpart of the Gabor formulation in (10)-(12) – by the convolution theorem, $\mathcal{F}^{-1}(U \odot M_{j,\theta}) = u * \mathcal{F}^{-1}(M_{j,\theta})$, and taking the modulus recovers the same phase-invariant energy feature. The mask $M_{j,\theta}$ is constructed as the product of a radial Gaussian bandpass term centered at frequency $f_j$ and a directional term selecting orientation $\theta$

$$M_{j,\theta}(f) = \exp\left(-\frac{(\parallel f \parallel - f_j)^2}{2\,(b_j/2)^2}\right) \cdot \frac{1 + \cos(2\pi f^{\top} e_{\theta})}{2} \tag{17}$$

Here the directional factor $[1 + \cos(2\pi f^\top e_\theta)]/2$ is a raised-cosine (Hann) modulation along $e_\theta = (\cos\theta, \sin\theta)^\top$, providing a soft, smoothly varying directional weighting that suppresses energy in the half-plane opposite to $e_\theta$; it does not produce a hard angular sector. The bandwidth is set to $b_j = 0.5\, f_j$. The center frequencies follow the geometric progression

$$f_j = f_{\min} \left(\frac{f_{\max}}{f_{\min}}\right)^{\frac{j}{J-1}}, \qquad j = 0, \dots, J-1 \tag{18}$$

with $f_{\min} = 0.02$ and $f_{\max} = 0.4$.

These are normalized spatial frequencies in cycles per resource-grid sample, as produced by the DFT frequency axis, where the Nyquist limit is 0.5; they carry no units in Hz. The value $f_{\min} = 0.02$ corresponds to a spatial period of $\approx 50$ grid points, capturing the large-scale energy envelope of the channel – wideband SNR, RMS delay spread, and the slow fade depth across the full slot. The value $f_{\max} = 0.4$ (near Nyquist) corresponds to a period of $\approx 2.5$ grid points, resolving fine-grained frequency-selective dips caused by closely-spaced multipath components. DC ($f = 0$) is excluded because it encodes only the mean signal power, which is not discriminative, while frequencies above 0.4 are dominated by noise rather than channel structure. The key design property is the constant quality factor $Q = f_j/b_j = 2$; because the bandwidth scales proportionally with center frequency ($b_j = 0.5\, f_j$), each filter occupies the same relative fraction of the frequency axis regardless of scale – the constant-$Q$ property of wavelet analysis. The geometric form of (18) encodes this design intent explicitly; extending to $J > 2$ scales would tile the frequency axis uniformly on a logarithmic grid. With the current $J = 2$ setting, the two filters cover complementary extremes – coarse channel quality and fine spectral nulls – with over four octaves of separation ($f_{\max}/f_{\min} = 20$).

**Two equivalent realizations**. Equation (15) can be evaluated exactly with cuFFT – useful as an accuracy reference – or, for deployment, the spatial kernel $\mathcal{F}^{-1}(M_{j,\theta})$ can be precomputed offline, fft-shifted, and cropped to a compact $5 \times 5$ support, after which the front-end reduces to two real-valued depthwise convolutions (one each for the real and imaginary parts) followed by the modulus, identical in form to (10)-(11). The cropped-kernel realization is fully compatible with ONNX export and TensorRT, while preserving the geometric scale design of (18). Interestingly, the $5 \times 5$ spatial truncation acts as a regularizer, discarding high-frequency noise artifacts that the exact FFT retains while preserving the discriminative multi-scale structure. Consequently, the cropped-kernel realization is not only deployment-friendly but also empirically superior for this task, as we observed much worse results when using the exact FFT scattering implementation.

**Why an FFT-based front-end**. Relative to the Gabor bank, the FFT scattering offers (i) geometric scale spacing, which provides a wider, more naturally logarithmic coverage of the delay/Doppler-induced frequency structure; (ii) explicit frequency-band control, since the passband is designed directly in the Fourier domain; and (iii) favor-able accelerator mapping – the precomputed filters are dispatched through general implicit-GEMM convolution kernels, which we find yield the lowest TensorRT kernel time (Sect. 4.5). These properties make the FFT front-end an attractive alternative to analytic Gabor kernels, particularly when precise control of the analyzed frequency bands is desired.

### 3.4 MCS Conditioning

In a realistic deployment, the gNB serves TBs spanning a range of modulation and coding schemes. Because each MCS index corresponds to a different code rate and operating SNR, the boundary separating decodable from non-decodable TBs shifts with MCS; a single threshold applied to channel-derived features therefore generalizes poorly across MCS values. We address this by conditioning the predictor on the MCS index $m$.

The scalar index is mapped through a learned embedding and a small projection,

$$e = \mathrm{ReLU}(W_e \, \mathrm{Embed}(m) + b_e) \tag{19}$$

and concatenated with the globally pooled scattering-CNN features $\phi = \mathrm{GAP}(F_\theta(z))$,

$$\tilde{\phi} = \mathrm{concat}(\phi, e) \tag{20}$$

before the final classification layer

$$\hat{p} = \sigma\left(w_o^\top \tilde{\phi} + b_o\right) \tag{21}$$

The embedding table supports up to 16 MCS indices, covering the 16-QAM set (MCS 5-10) used here with headroom for extension. The conditioning branch adds only a few hundred parameters yet, as shown in Sect. 4.3, allows the model to learn MCS-dependent decision boundaries and markedly reduces the false-alarm rate of the scattering predictors.

### 3.5 Neural Classification Head

The scattering tensor $z$ is processed by a compact CNN classifier $F_\theta(\cdot)$. A $1 \times 1$ convolution first performs a channel projection that mixes and reweights the responses of the filter bank at each grid location while preserving spatial resolution. A standard $3 \times 3$ convolution with batch normalization and ReLU then aggregates local time-frequency context. The representation is refined by a sequence of depthwise-separable convolutions (each with batch normalization and ReLU) arranged in an expand–contract pattern, which provides representational capacity at low computational cost. Global average pooling collapses the grid into a feature vector $\phi$, which is mapped to the CRC probability by a sigmoid output as in (4) – or, when MCS conditioning is used, fused with the MCS embedding via (19)-(21). The same backbone is used for all front-ends, isolating the effect of the feature extractor and of MCS conditioning.

### 3.6 Evidential Learning for Uncertainty Quantification

For URLLC applications where minimizing missed detections is paramount, we extend the predictor with an evidential deep learning formulation that quantifies prediction uncertainty in a single forward pass, following the formulation in [10]. Unlike sampling-based Bayesian methods such as MC Dropout, which require multiple forward passes and increase latency, evidential deep learning predicts a Dirichlet distribution over class probabilities, enabling uncertainty decomposition without additional inference overhead. The model outputs Dirichlet concentration parameters $\boldsymbol{\alpha} = [\alpha_0, \alpha_1]$, where $\alpha_k > 0$ for each class $k \in \{0,1\}$, with $k = 0$ denoting CRC fail and $k = 1$ denoting CRC pass. The final dense layer applies a softplus activation $\alpha_k = \log(1 + e^{z_k})$, to its logits $z_k$, ensuring the concentration parameters are strictly positive as required by the Dirichlet distribution. This replaces the sigmoid output of the non-evidential predictor. The class probabilities are obtained by normalization –

$$\mathbf{p} = \frac{\boldsymbol{\alpha}}{\sum_k \alpha_k} \tag{22}$$

where $p_k$ represents the predicted probability of class $k$. The one-hot label vector is $\mathbf{y} = [y_0, y_1]$ with $y_1 = c$ and $y_0 = 1 - c$. The evidence for each class is $e_k = \alpha_k - 1 \geq 0$ and the Dirichlet strength is $S = \sum_k \alpha_k$.

The evidential loss function combines a classification risk term with a KL divergence regularization

$$\mathcal{L}_{\text{evi}} = \mathcal{L}_{\text{risk}} + \lambda \cdot \text{KL}\big(\text{Dirichlet}(\tilde{\boldsymbol{\alpha}}) \parallel \text{Dirichlet}(\mathbf{1})\big) \tag{23}$$

where $\lambda$ controls the regularization strength. The classification risk is the expected log-loss under the Dirichlet, using the identity $\mathbb{E}_{p\sim\text{Dir}(\boldsymbol{\alpha})}[\log p_k] = \psi(\alpha_k) - \psi(S)$

$$\mathcal{L}_{\text{risk}} = \sum_k y_k \big(\psi(S) - \psi(\alpha_k)\big) \tag{24}$$

To avoid penalizing the model for accumulating evidence on the correct class, the KL term operates on a masked concentration vector $\tilde{\alpha}_k = y_k + (1 - y_k)\alpha_k$, which reduces the ground-truth class concentration to 1 (zero additional evidence) while retaining evidence for incorrect classes. The KL divergence between this masked Dirichlet and a uniform prior is

$$\text{KL}\big(\text{Dirichlet}(\tilde{\boldsymbol{\alpha}}) \parallel \text{Dirichlet}(\mathbf{1})\big) = \log \frac{\Gamma(\tilde{S})}{\Gamma(K)\prod_k \Gamma(\tilde{\alpha}_k)} + \sum_k (\tilde{\alpha}_k - 1)[\psi(\tilde{\alpha}_k) - \psi(\tilde{S})] \tag{25}$$

where $\tilde{S} = \sum_k \tilde{\alpha}_k$, $\Gamma(\cdot)$ is the gamma function, $\psi(\cdot)$ is the digamma function, and $K = 2$.

A key advantage of the evidential formulation is the decomposition of total uncertainty into aleatoric and epistemic components. For classification with a Dirichlet prior, we define:

- Total uncertainty: entropy of the predictive distribution

$$u_{\text{total}} = -\sum_k p_k \log p_k \tag{26}$$

- Aleatoric uncertainty: expected entropy of the categorical distribution under the Dirichlet

$$u_{\text{aleatoric}} = \sum_k \frac{\alpha_k}{S}\big(\psi(S+1) - \psi(\alpha_k + 1)\big) \tag{27}$$

- Epistemic uncertainty: the difference, representing model uncertainty reducible with more data

$$u_{\text{epistemic}} = u_{\text{total}} - u_{\text{aleatoric}} \tag{28}$$

At inference, we employ a conservative decision rule that predicts failure when epistemic uncertainty exceeds a threshold $\tau_u$

$$\hat{c}_{\text{cons}} = \mathbb{1}\big[u_{\text{epistemic}} > \tau_u\big] \cdot 0 + \mathbb{1}\big[u_{\text{epistemic}} \le \tau_u\big] \cdot \mathbb{1}\big[p_{\text{pass}} \ge \tau\big] \tag{29}$$

where $p_{\text{pass}} \equiv p_1 = \alpha_1 / S$ is the predicted CRC-pass probability from (22), and $\tau$ is the standard decision threshold (typically 0.5). This rule biases the predictor toward catching failures when the model is uncertain, which is particularly valuable for URLLC where missed detections are safety-critical. The evidential formulation maintains single-pass inference (low latency) while providing actionable uncertainty estimates, unlike MC Dropout which requires multiple stochastic forward passes.

# 4 Experimental Results

This section evaluates the proposed predictors against a pure-CNN baseline and analyzes classification performance, the effect of MCS conditioning, model complexity, and inference latency.

## 4.1 Simulation Setup

All simulations were conducted using NVIDIA Aerial CUDA-Accelerated RAN [11], which provides GPU-accelerated implementation of 5G NR physical layer processing. The experiments were implemented using the Python interface of the framework, PyAerial. Training data generation and receiver evaluation were performed using the PUSCH example pipelines provided by the framework. These pipelines were used both to generate training data – uplink I/Q, LS channel estimates, and the ground-truth CRC outcome from the LDPC decoder, and to execute the inference pipeline using TensorRT engines.

To stress generalization, the training set spans a range of channel conditions and six MCS indices, while the held-out test set additionally includes channel configurations (delay-spread and UE-speed combinations) not present in training, yielding an out-of-distribution evaluation. The main simulation parameters used in the experiments are summarized in Table 2, with each slot generated using channel parameters sampled from a uniform distribution.

The predictor is trained against the true CRC labels using the BCE loss (5). The dataset is class-imbalanced (approximately 74% CRC-fail, 26% CRC-pass), reflecting the challenging channel conditions in the simulation sweep; inverse-frequency class weights are applied during training to prevent the classifier from defaulting to the majority class. The experiments were conducted on an NVIDIA L4 GPU installed in a Dell PowerEdge XR8620t platform, which represents an edge deployment configuration suitable for AI-accelerated RAN workloads.

**Table 2.** Simulation parameters

| Parameter | Value |
|---|---|
| Training samples | 30000 slots |
| Test samples | 20000 slots |
| Channel model | TDL/CDL-A, B, C |
| Delay Spread | 0-300 ns |
| UE speed (training) | 3-110 kmph |
| UE speed (inference) | 3-140 kmph |
| UE antennas | 1 |
| gNB antennas | 2 |
| Transmission Layers | 1 |
| DMRS symbols | 1 |
| PUSCH symbols | 13 |
| PRBs | 273 |
| SCS | 30 kHz |
| MCS Table | 256-QAM |
| MCS Index | 5-10 (16-QAM) |

### 4.2 Classification Performance

We compare six configurations sharing the same CNN backbone – a pure CNN baseline and the Gabor+CNN and FFT+CNN scattering predictors, each evaluated without and with MCS conditioning. Performance on the 20,000-slot out-of-distribution test set is reported in Fig. 4 using accuracy, MDR, and FAR; the label convention is $c=0$ (CRC fail) and $c=1$ (CRC pass). Two findings stand out. First, the scattering front-ends provide essential inductive bias – replacing the pure CNN with either Gabor or FFT scattering raises accuracy from 82% to 90% and cuts MDR from 19.20% to below 4.5%. The pure CNN, lacking a spectral prior, tends to overpredict “pass” on unseen channels, resulting in an MDR that is unacceptable for URLLC (roughly one in five corrupted TBs passing silently). Second, the FFT front-end exhibits a small but consistent MDR advantage over Gabor (3.77% vs. 4.43%) in the absence of MCS conditioning, attributable to its geometric scale spacing.

### 4.3 Effect of MCS Conditioning

Without MCS, the scattering predictors exhibit low MDR but high FAR (25-26%), because a single decision boundary must serve six different code rates simultaneously. Adding the MCS embedding of (19)-(21) lets the model learn MCS-dependent boundaries and sharply reduces FAR – for FFT+CNN the FAR drops from 26.21% to 10.00% and accuracy rises to 95%, while for Gabor+CNN the FAR drops from 25.90% to 18.28% with MDR improving to 2.90%. In contrast, although MCS conditioning yields a gain for the pure CNN for MDR (19.20% to 11.95%), it increases its FAR because the backbone features are too weak for the MCS context to exploit. MCS conditioning therefore synergizes specifically with the scattering front-ends – the front-end captures how much spectral distortion is present, while the MCS index informs the model how much distortion is tolerable at the operating code rate. The best overall operating point is FFT+CNN+MCS (95% accuracy, 3.91% MDR, 10.00% FAR); Gabor+CNN+MCS offers the lowest MDR (2.90%) at the cost of higher FAR, a useful alternative when missed detections must be minimized.

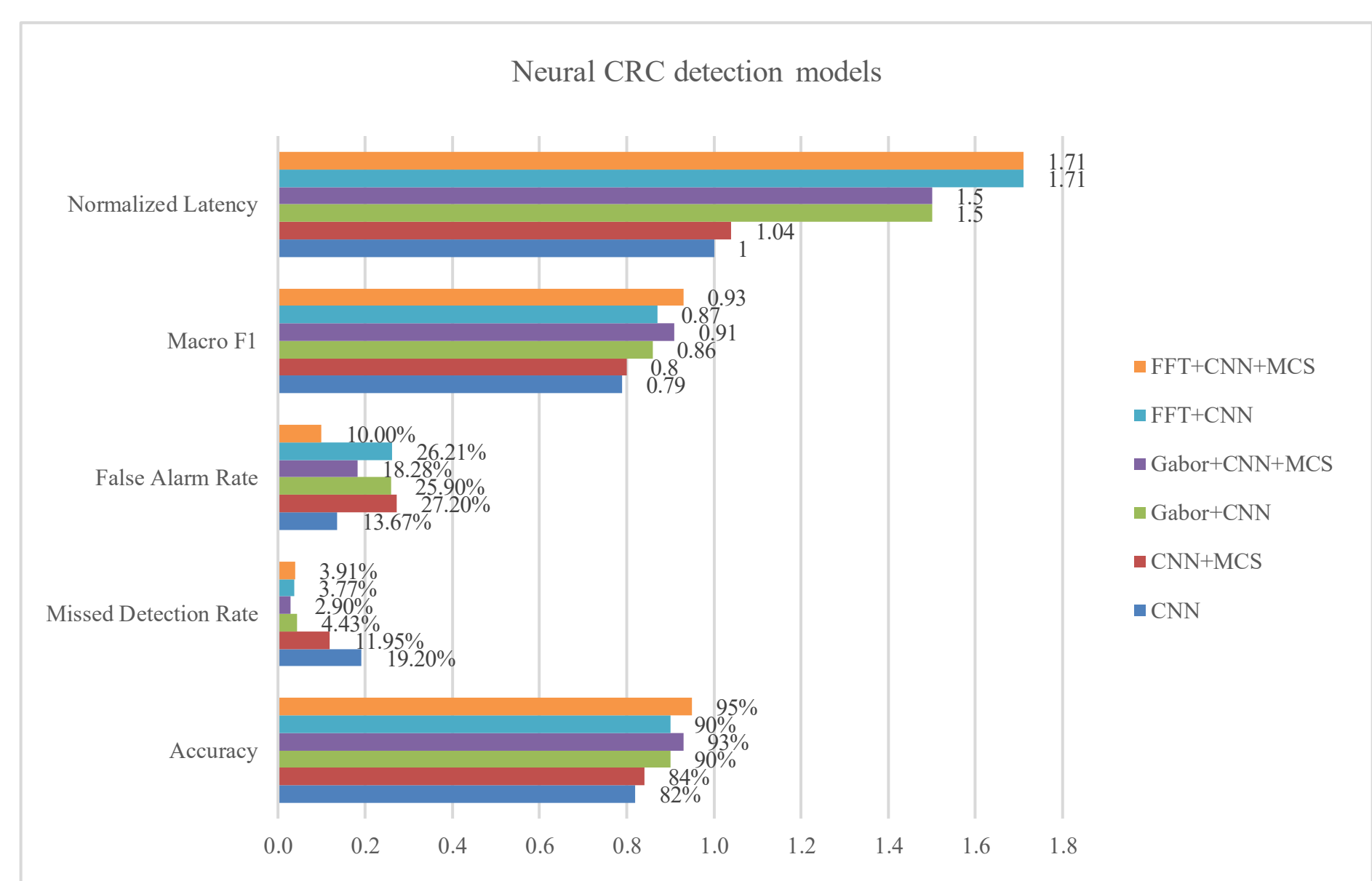

**Fig. 4.** Classification performance on test set

Overall, the comparison reveals a consistent tradeoff between the two front-ends. The FFT-based design provides the strongest overall performance, achieving higher accuracy and lower FAR, while the Gabor-based design achieves the lowest MDR, making it suitable for deployments that prioritize minimizing missed detections. Both front-ends exhibit comparable computational cost, and as discussed in Sect. 4.5, the FFT-based implementation also benefits from more efficient accelerator execution.

We further note that the decision threshold $\tau$ in (6) provides a deployment-time control to bias the predictor toward catching failures (lower MDR) at the expense of additional false alarms – a favorable trade for URLLC, since a false alarm costs a single retransmission whereas blind repetition retransmits every TB 2–8×.

### 4.4 Model Complexity

A central motivation for incorporating scattering features is to reduce the size of the learnable detector. The compact backbone used here contains approximately 7,385 trainable parameters. The Gabor front-end adds 3,200 fixed (non-trainable) filter weights, and the FFT front-end adds zero trainable parameters, as its filters are graph constants; MCS conditioning adds roughly 200 parameters. It is important to distinguish parameter count from computational cost. The FFT front-end, despite having no trainable parameters, does add runtime FLOPs through its $J \times L = 8$ pairs of depth-wise convolutions – each with kernels derived from the precomputed spatial filters. This compute overhead is visible in the TensorRT profiling results of Sect. 4.5, where Gabor and FFT front-ends add comparable inference time (approximately 0.29-0.30 ms real compute) relative to the plain CNN (0.11 ms).

To investigate whether additional learnable capacity could compensate for the absence of a structured front-end, we evaluated a scaled-up pure-CNN backbone with the hidden dimension doubled from 8 to 16 channels. The accuracy improvement was marginal while inference latency increased significantly, confirming that the performance bottleneck of the plain CNN is not parameter count but the lack of a signal-processing-informed feature extraction stage. By contrast, moving the multi-scale time-frequency analysis into a fixed front-end keeps the learnable classifier compact while improving robustness on unseen channels.

### 4.5 Inference Latency

Models are exported to TensorRT (FP16, static shapes) and profiled with NVIDIA Nsight Systems. Fig. 5 reports the GPU kernel execution time per inference (batch size 1) measured on an NVIDIA L4 GPU, separating the total time from the dominant kernel conversion overhead – an I/O-format artifact that can be eliminated by aligning the engine's input layout to its internal NHWC compute path. MCS conditioning adds negligible compute and is not shown separately.

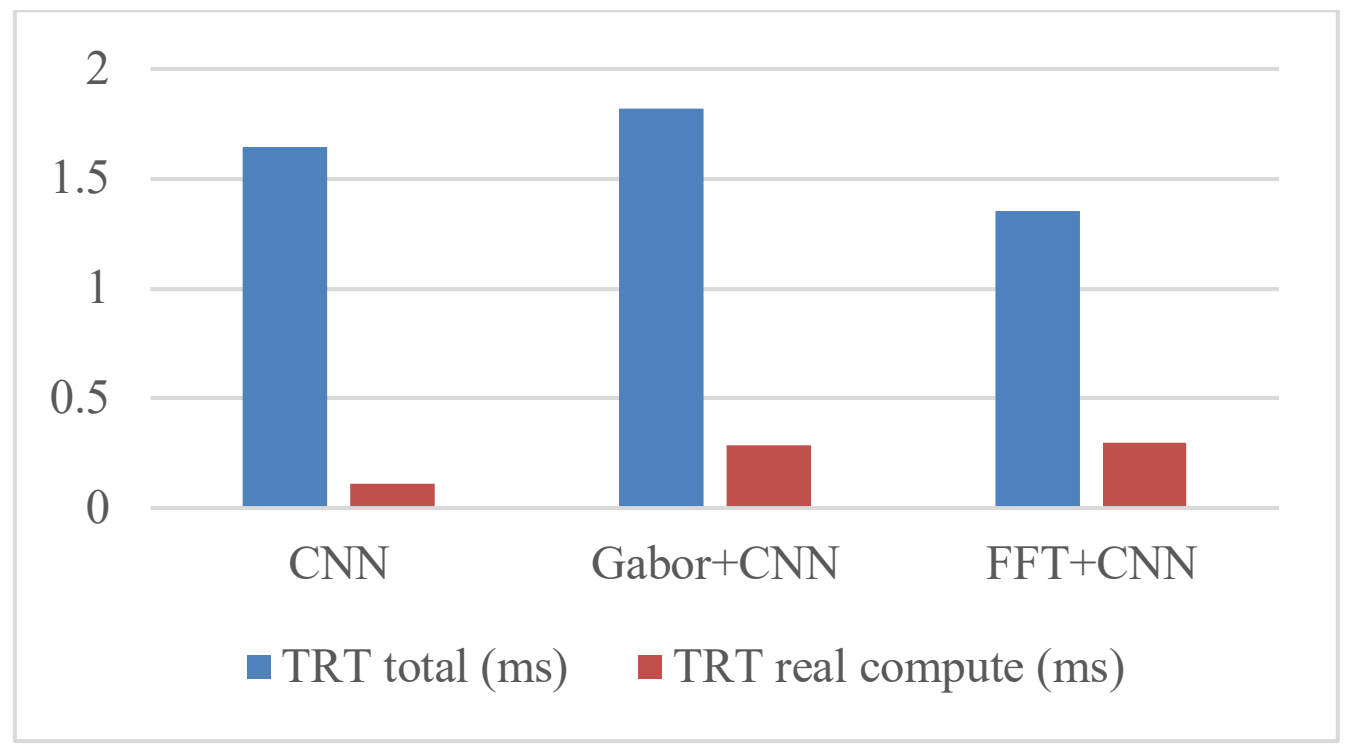

**Fig. 5.** Inference latency comparison

The FFT front-end achieves the lowest total GPU time (1.353 ms) owing to a smaller kernel conversion share; its real model compute (0.299 ms) is comparable to Gabor front-end (0.287 ms). The difference in kernel routing accounts for this, since the FFT filters stored as graph constants are dispatched through general implicit-GEMM convolutions, while the Gabor kernel depthwise path incurs a larger I/O-format conversion. After eliminating the kernel conversion overhead, all three models converge to 0.11-0.30 ms of real compute, with the plain CNN at 0.11 ms (no front-end) and both scattering models at approximately 0.29-0.30 ms.

It is important to note that these measurements are obtained with the full 100 MHz allocation of 273 PRBs, representing an eMBB-grade resource grid. This configuration is directly applicable to eMBB use cases and establishes an upper bound on the predictor's input size. For URLLC deployments, transport blocks are typically scheduled on a much narrower bandwidth – often 25–50 PRBs or fewer – and the resource grid presented to the network is correspondingly smaller. Since the inference time scales with input size, the real model compute is expected to be proportionally shorter for URLLC TBs, further reinforcing the suitability of the proposed predictor for latency-sensitive deployments.

### 4.6 Evidential Learning for High-reliability URLLC

For URLLC use cases where minimizing missed detections is the primary objective, we evaluate the evidential deep learning formulation described in Sect. 3.6 as an additional option that can be combined with either the FFT or Gabor scattering front-ends. The evidential model provides uncertainty quantification in a single forward pass, enabling a conservative decision rule that predicts failure when epistemic uncertainty exceeds a threshold.

We trained an FFT+CNN+MCS predictor with the evidential output head (Dirichlet parameters) using KL weight $\lambda = 0.01$. The results on the out-of-distribution test set are summarized in Table 3. The evidential formulation achieves the lowest MDR among all evaluated models (0.76%) when using the conservative decision rule with epistemic uncertainty threshold $\tau_u = 0.025$. This represents an 81% reduction in MDR compared to the baseline FFT+CNN+MCS predictor (3.91%), at the cost of increased FAR (42.24% vs 10.00%). The standard evidential decision (without uncertainty threshold) yields MDR 3.90% and FAR 8.86%, which already matches the baseline MDR while improving FAR by 1.1 percentage points. The uncertainty decomposition shows that 19.3% of test samples are flagged as uncertain at threshold 0.025, with a mean epistemic uncertainty of 0.030 among uncertain samples. The conservative rule correctly identifies 46.2% of uncertain samples as true failures, demonstrating a meaningful correlation between epistemic uncertainty and prediction difficulty. The aleatoric uncertainty averages 0.204 across all samples, reflecting inherent channel ambiguity.

**Key observations**. The conservative decision rule significantly reduces MDR (0.76%) at an acceptable FAR increase for safety-critical URLLC deployments. This tradeoff is justified by the asymmetric cost structure of URLLC, where missed detections are far more detrimental than unnecessary retransmissions, and even at 42.24% FAR, the expected retransmission overhead is approximately 0.42 per transport block, which remains substantially lower than blind repetition schemes that require 1–7 additional transmissions per transport block (for 2–8× repetition). The uncertainty threshold $\tau_u$ provides a deployment-time knob to trade off MDR vs FAR – lower thresholds increase uncertainty flagging and further reduce MDR, while higher thresholds reduce FAR. The evidential formulation is compatible with both FFT and Gabor front-ends, offering a flexible enhancement for high-reliability use cases.

Under the standard decision rule, the evidential model matches the non-evidential FFT+CNN+MCS predictor in accuracy and MDR, while achieving a lower FAR (8.86% vs 10.00%). This improvement stems from the KL divergence regularization in the evidential loss, which penalizes misleading evidence for incorrect classes and acts as an implicit false-alarm suppressor.

**Table 3.** Evidential Deep Learning Performance (FFT+CNN+MCS, KL weight = 0.01)

| Decision rule | Accuracy | MDR | FAR | Uncertain % |
|---|---|---|---|---|
| Standard ($p_{\text{pass}} \geq 0.5$) | 95% | 3.90% | 8.86% | — |
| Conservative ($u_{\text{epistemic}} > \tau_u = 0.025$) | 89% | 0.76% | 42.24% | 19.30% |

The evidential model maintains single-pass inference (low latency) while providing actionable uncertainty estimates. For URLLC applications where missed detections are unacceptable, the evidential predictor with conservative decision provides the best safety margin. For eMBB and other bandwidth-efficient operating modes, the standard decision rule provides excellent overall accuracy with low MDR and FAR. This flexibility allows a single evidential predictor to adapt to differing reliability and resource-efficiency requirements without retraining.

## 5 Conclusion and Future Work

We presented a hybrid neural CRC predictor for the 5G NR PUSCH that enables early, low-latency link-adaptation decisions for URLLC. Borrowing the input convention of neural receivers – received I/Q samples conditioned on LS channel estimates – the predictor estimates the post-decoding CRC outcome directly from the received resource grid, well ahead of the conventional decoding chain. A fixed first-order scattering front-end supplies multi-scale, phase-invariant time-frequency features to a compact CNN classifier, and we studied two complementary front-end realizations – a Gabor wavelet bank in the spatial domain and an FFT-based bank with geometric scale spacing in the frequency domain.

On an out-of-distribution, multi-MCS test set, both scattering front-ends substantially outperform a pure CNN, reducing the safety-critical missed-detection rate from 19.20% to below 4.5%. Introducing the MCS index as an auxiliary input resolves the elevated false-alarm rate of the scattering models operating across multiple code rates. The FFT-based predictor with MCS conditioning achieves the best overall operating point at 95% accuracy, 3.91% MDR, and 10.00% FAR, while the Gabor-based variant achieves the lowest MDR at 2.90%. Both front-ends are TensorRT-compatible and execute within the slot processing budget on an edge-class GPU, with the FFT front-end attaining the lowest measured kernel time.

These results indicate that early CRC prediction is a practical mechanism for reducing the effective HARQ round-trip time and avoiding wasteful blind repetitions in URLLC, restoring deterministic latency for mission-critical traffic. For applications requiring the highest reliability, we further demonstrated an evidential deep learning formulation that quantifies prediction uncertainty in a single forward pass while achieving 0.76% MDR. Importantly, the same trained evidential model can support both eMBB and URLLC deployments by selecting different decision rules at inference time. Future work includes fusing the models into the PUSCH pipeline for further latency reduction, integration with the HARQ feedback loop, and extending to additional modulation orders and DMRS configurations for unified operation across the full link-adaptation range.

**Acknowledgements**. We acknowledge the use of approved Generative AI tools to support code refinement, document preparation, and editorial review. These tools were used solely to assist in the implementation and presentation and did not contribute to the generation of original research ideas, experimental design, or conclusions. All core contributions presented in this paper are the author's own.